\documentclass[12pt]{article}
\usepackage{graphicx}
\begin{document}
\centerline{\bf Potts model with $q$  states on directed Barab\'asi-Albert networks}
 
\bigskip
\centerline{F.W.S. Lima}
 
\bigskip
\noindent
Departamento de F\'{\i}sica, 
Universidade Federal do Piau\'{\i}, 57072-970 Teresina - PI, Brazil
 
\medskip
  e-mail: wel@ufpi.br
\bigskip
 
{\small Abstract: On directed Barab\'asi-Albert networks with two and
 seven  neighbours selected by each added site, the Ising model with
 spin $S=1/2$ was seen
 not to show a spontaneous magnetisation. Instead, the decay time for 
 flipping of the magnetisation followed an Arrhenius law for Metropolis
 and Glauber algorithms, but for Wolff cluster flipping the
 magnetisation  decayed exponentially with time.  However, on these
 networks the Ising model spin $S=1$ was seen to show a spontaneous magnetisation.
 In this model with spin $S=1$ a first-order phase
 transition for values of connectivity $z=2$ and $z=7$ 
 is well defined. On these same networks the
 Potts model with $q=3$ and $8$ states is now studied through Monte Carlo simulations.
We have obtained also for $q=3$ and $8$ states  a first-order phase
 transition for values of connectivity $z=2$ and $z=7$ of the 
 directed Barab\'asi-Albert network. Theses results are different from the results
 obtained for same model on two-dimensional lattices, where for $q=3$ the phase transition
 is of second order, while for $q=8$ the phase transition is first-order.}
 
 
\bigskip

 {\bf Introduction}
 
 Sumour and Shabat \cite{sumour,sumourss} investigated Ising models with
 spin $S=1/2$ on directed Barab\'asi-Albert networks \cite{ba} with
 the usual Glauber dynamics.  No spontaneous magnetisation was 
 found, in contrast to the case of undirected  Barab\'asi-Albert networks
 \cite{alex,indekeu,bianconi} where a spontaneous magnetisation was
 found lower a critical temperature which increases logarithmically with
 system size. In S=1/2 systems on undirected, scale-free hierarchical-lattice
small-world networks \cite{nihat}, conventional and algebraic
(Berezinskii-Kosterlitz-Thouless) ordering, with finite transition
temperatures, have been found. Lima and Stauffer \cite{lima} simulated
 directed square, cubic and hypercubic lattices in two to five dimensions
 with heat bath dynamics in order to separate the network effects  form
 the effects of directedness. They also compared different spin flip
 algorithms, including cluster flips \cite{wang}, for
 Ising-Barab\'asi-Albert networks. They found a freezing-in of the 
 magnetisation similar to  \cite{sumour,sumourss}, following an Arrhenius
 law at least in low dimensions. This lack of a spontaneous magnetisation
 (in the usual sense)
 is consistent with the fact
 that if on a directed lattice a spin $S_j$ influences spin $S_i$, then
 spin $S_i$ in turn does not influence $S_j$, 
 and there may be no well-defined total energy. Thus, they show that for
 the same  scale-free networks, different algorithms give different
 results. More recently, Lima \cite{lima1} simulated the the Ising model
 for spin $S=1$ on directed 
 Barab\'asi-Albert network and different from the Ising model for
 spin $S=1/2$, the order-disorder phase transition of 
 order parameter is well defined in this system. He has 
 obtained a first-order phase transition for values of 
 connectivity $z=2$ and $z=7$ of the directed Barab\'asi-Albert network.
 Now we study the Potts model for  $q=3$ and $8$ states on directed 
 Barab\'asi-Albert network for values of 
 connectivity $z=2$ and $z=7$.  and different from the Ising model for
 spin $S=1/2$, the order-disorder phase transition of 
 order parameter is well defined in this system. We have 
 obtained a first-order phase transition for values of 
 connectivity $z=2$ and $z=7$ of the directed Barab\'asi-Albert network.
\bigskip
 
\begin{figure}[hbt]
\begin{center}
\includegraphics[angle=-90,scale=0.46]{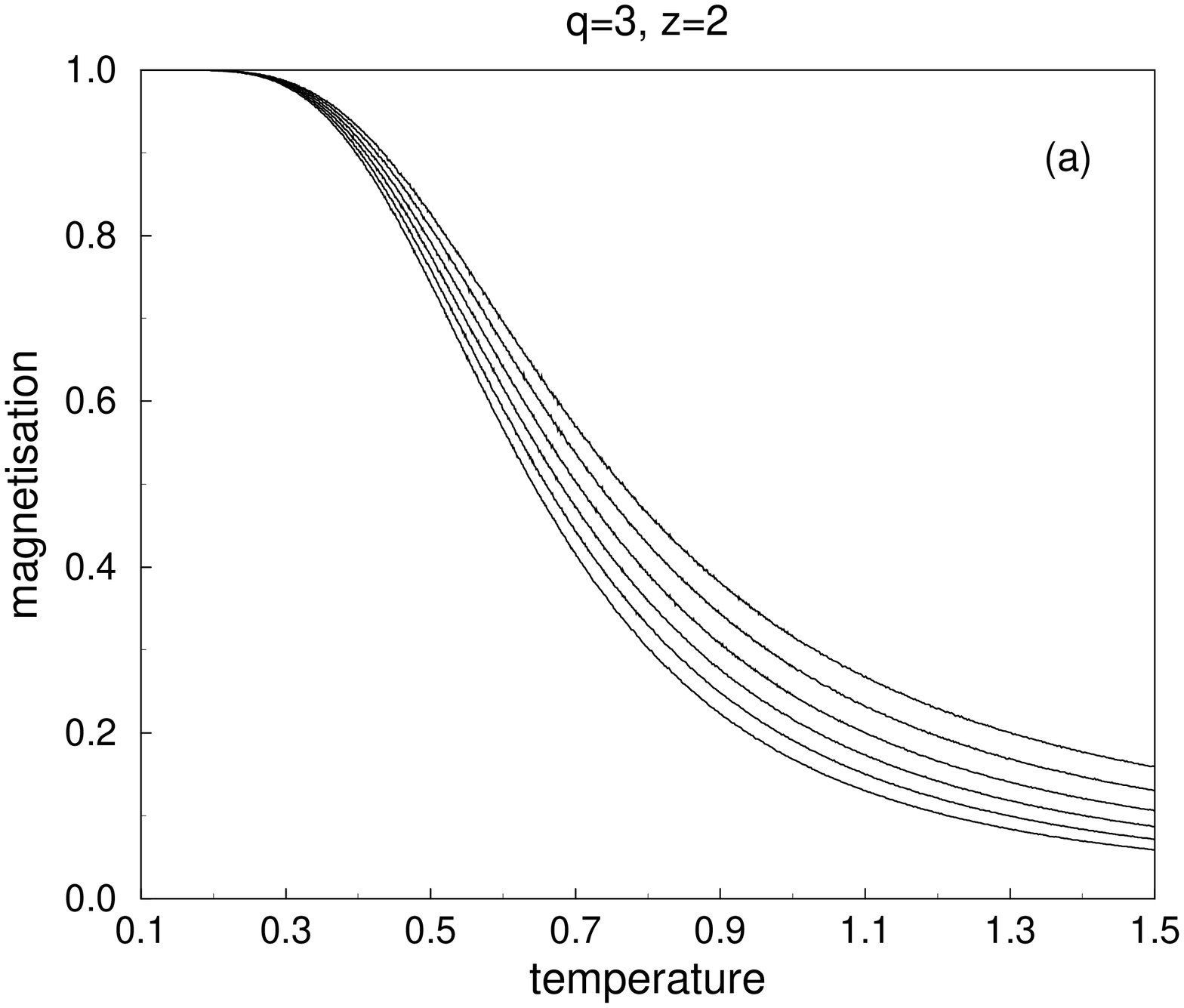}
\includegraphics[angle=-90,scale=0.46]{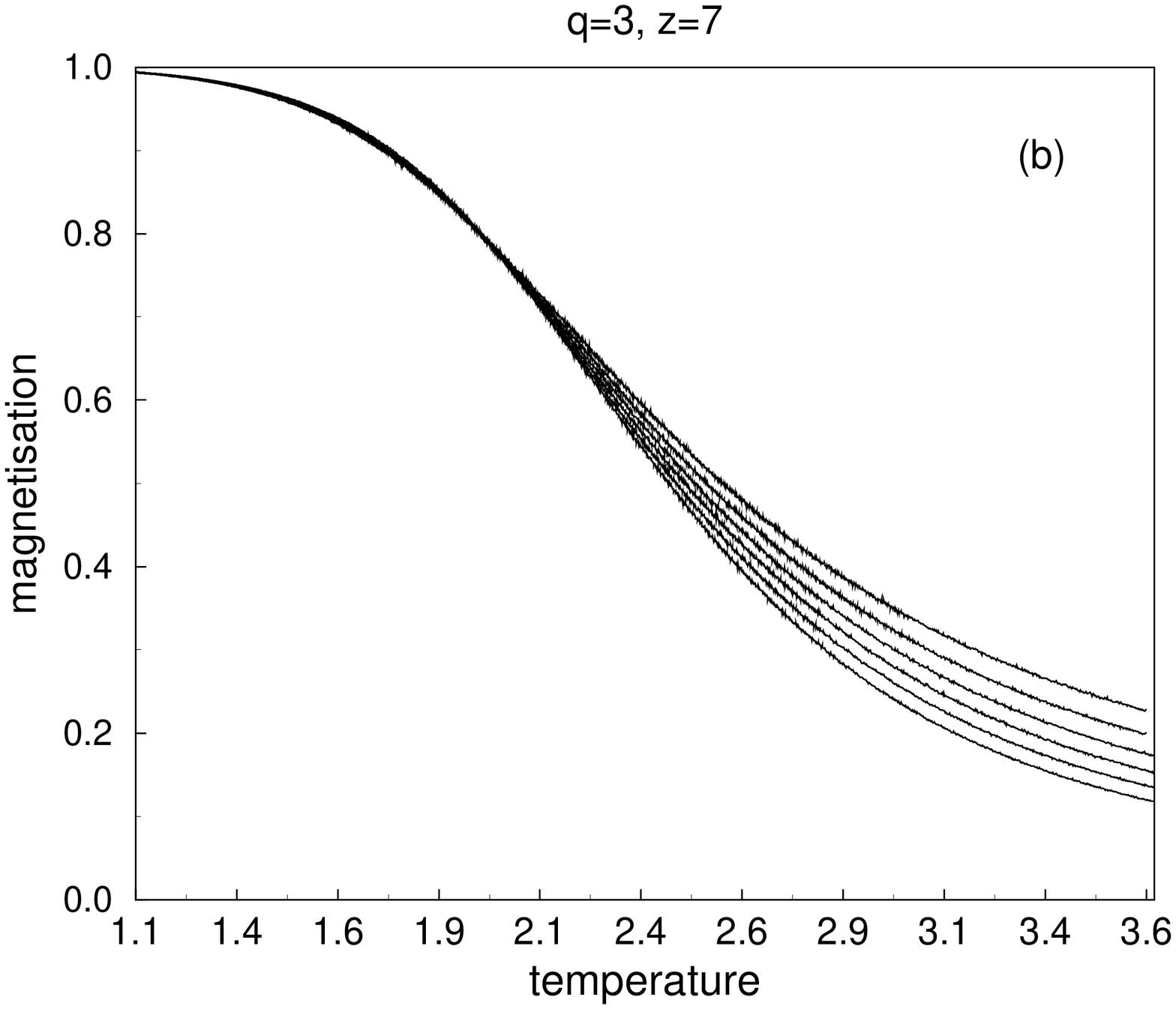}
\end{center}
\caption{
Plot of the magnetisation $M$ versus $K$ for $q=3$ and severals systems sizes ($N=250$, $500$, $1000$, $2000$, $4000$, and $8000$ ). In  part (a) $z=2$ and part (b) $z=7$.}
\end{figure}
  
\bigskip
 
\begin{figure}[hbt]
\begin{center}
\includegraphics[angle=-90,scale=0.46]{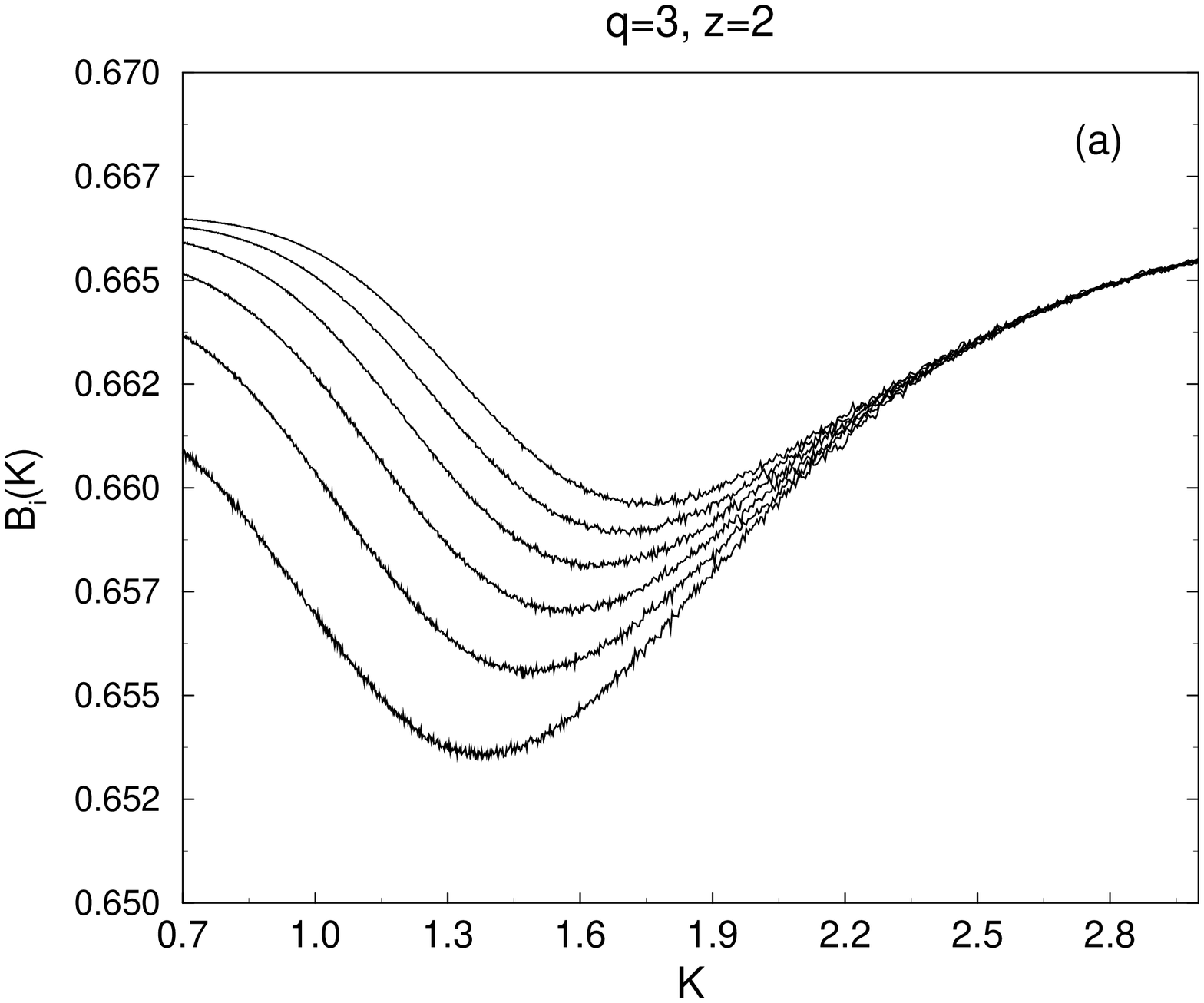}
\includegraphics[angle=-90,scale=0.46]{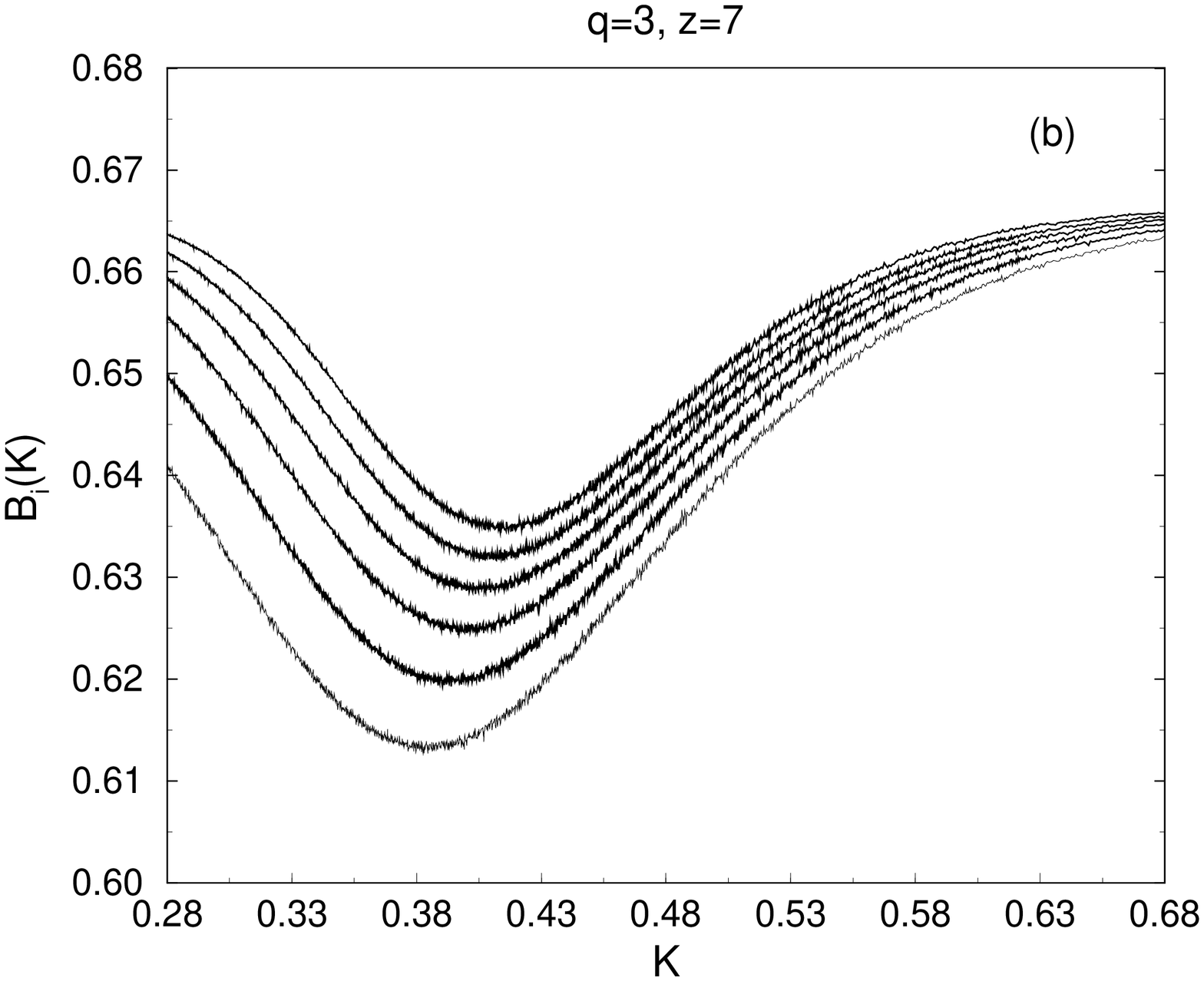}
\end{center}
\caption{
Plot of the Binder cumulant $B_{i}(K)$ versus $K$ for $q=3$ and severals systems sizes ($N=250$, $500$, $1000$, $2000$, $4000$, and $8000$ ). In part (a) $z=2$ and part (b) $z=7$.}
\end{figure}
 
\begin{figure}[hbt]
\begin{center}
\includegraphics[angle=-90,scale=0.46]{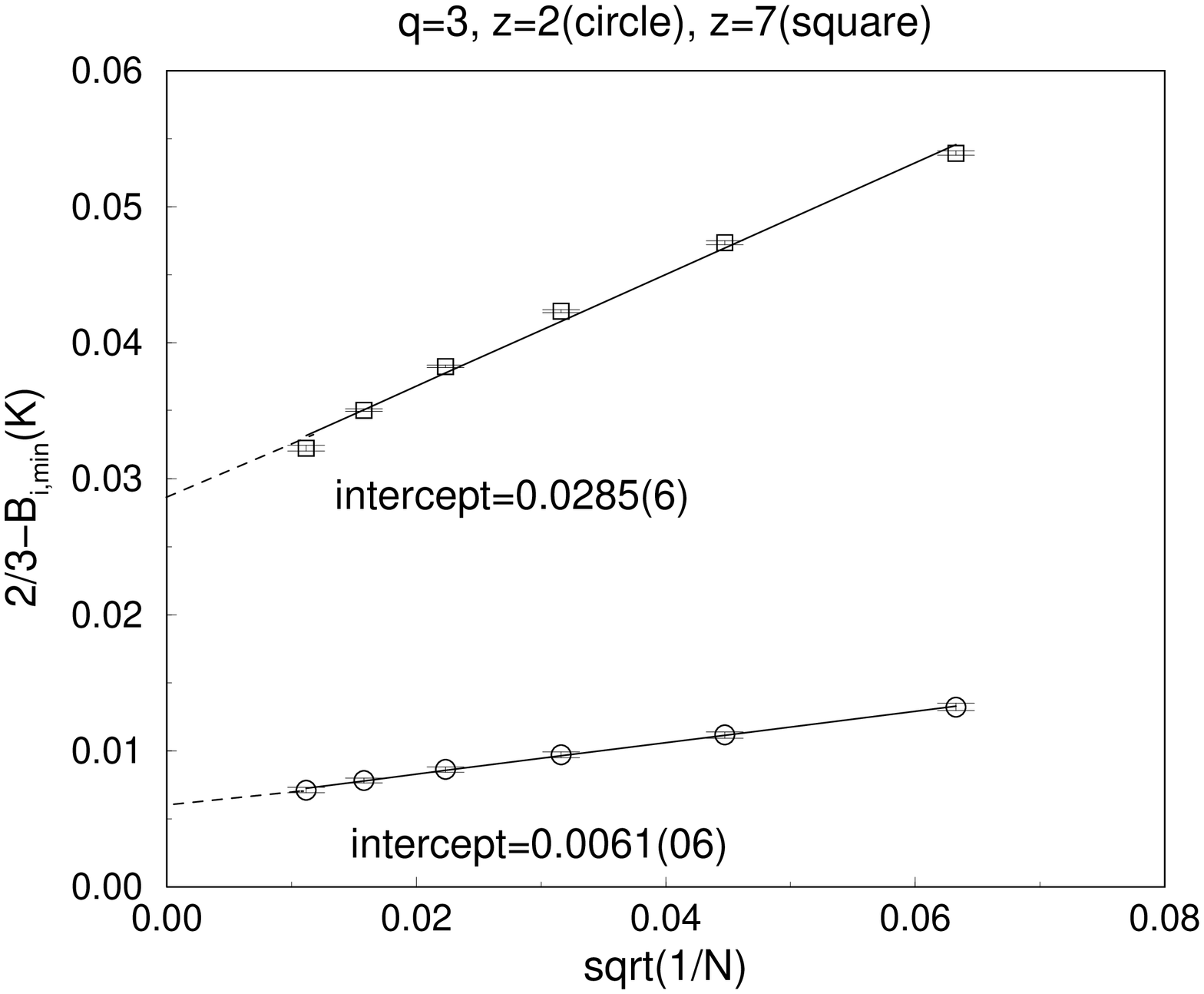}
\includegraphics[angle=-90,scale=0.46]{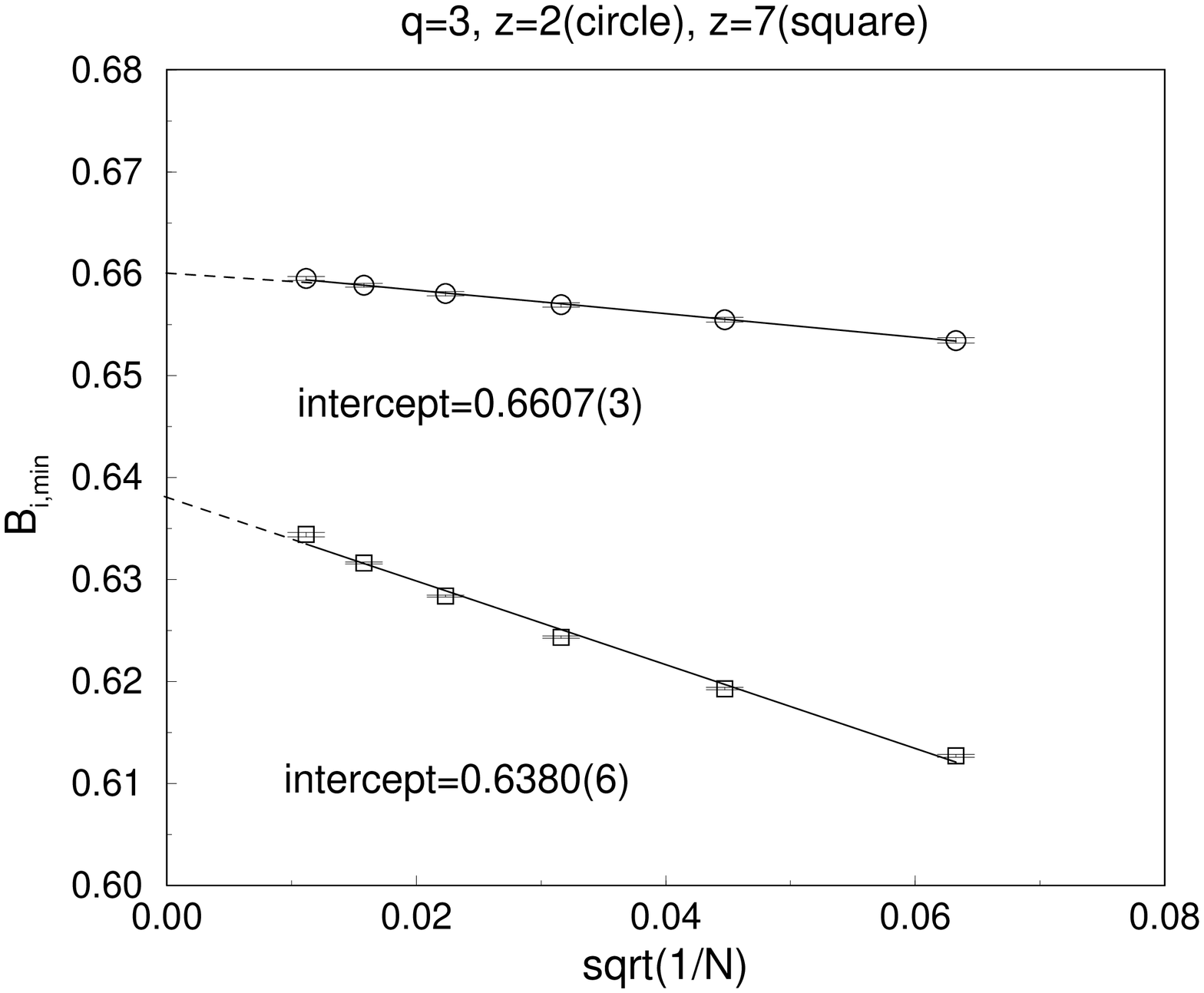}
\end{center}
\caption{Plot of the Binder parameter $2/3-B_{i}(K)$ versus $1/N$ for $z=2$ (circle) and $z=7$ (square),part (a) and  $B_{i,min}(K)$ versus $1/N$, part (b) for $q=3$.
} 
\end{figure}
 
\bigskip

\begin{figure}[hbt]
\begin{center}
\includegraphics[angle=-90,scale=0.46]{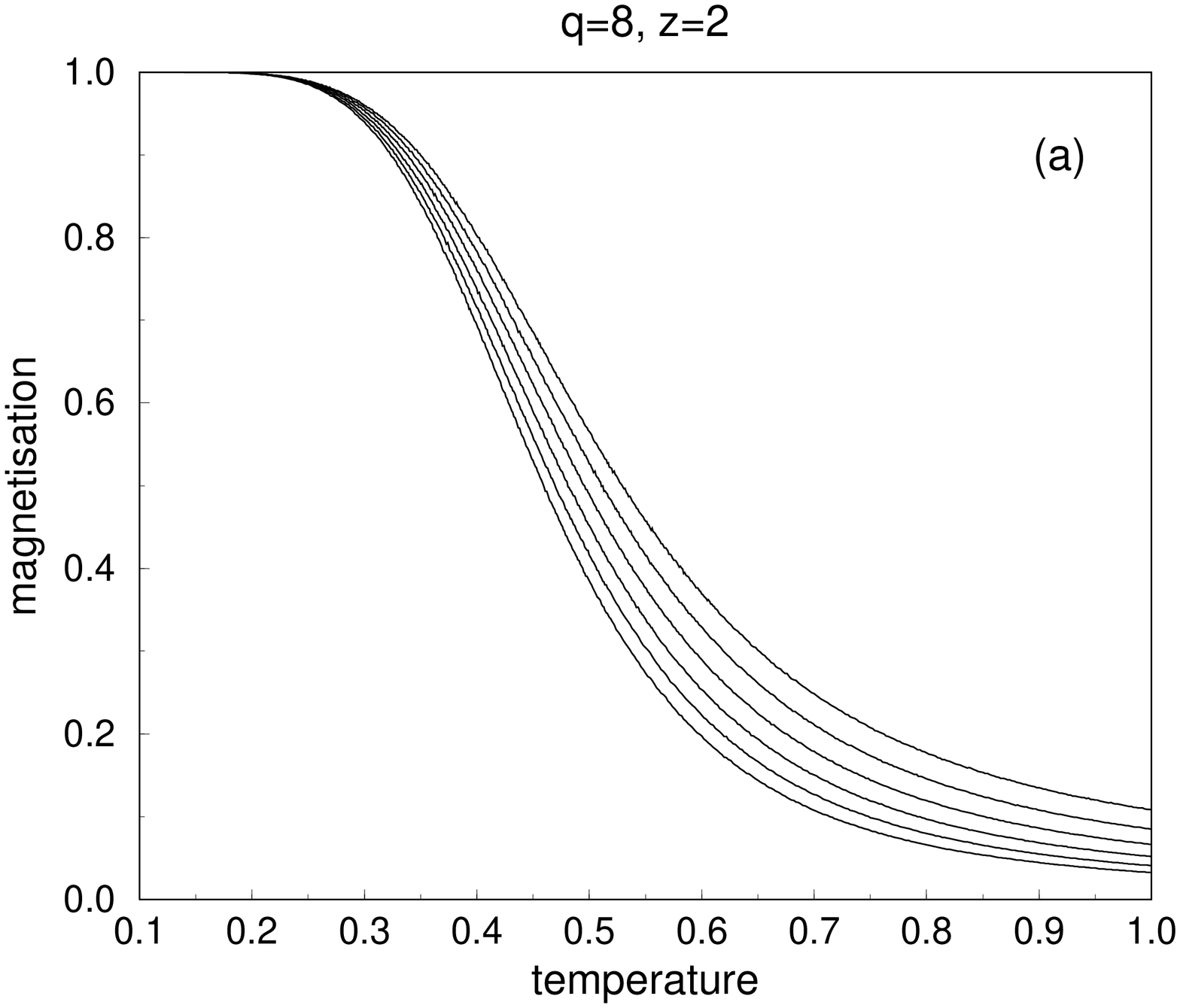}
\includegraphics[angle=-90,scale=0.46]{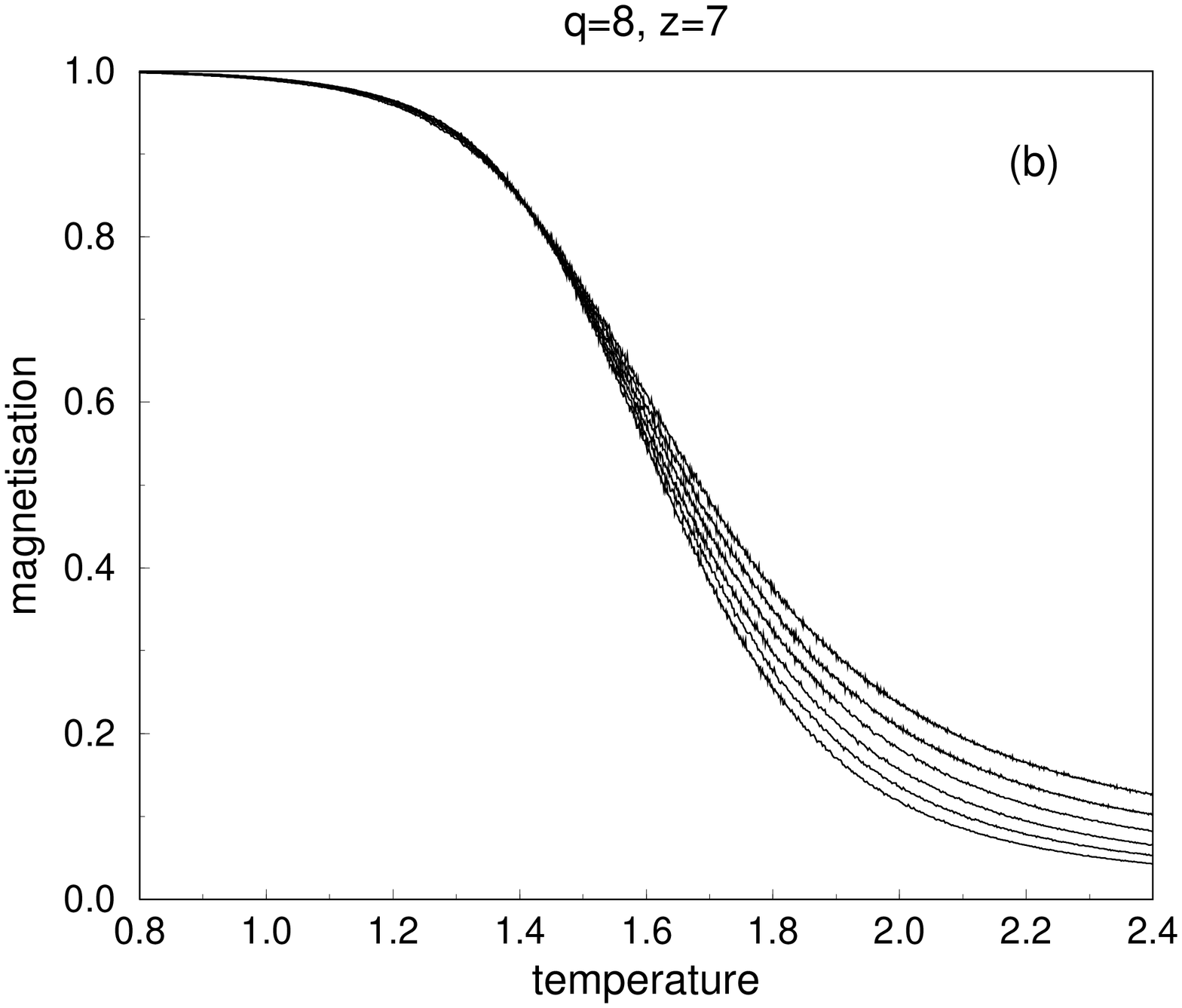}
\end{center}
\caption{Plot of the magnetisation $M$ versus $K$ for $q=8$ and severals systems sizes ($N=250$, $500$, $1000$, $2000$, $4000$, and $8000$ ). In part (a) $z=2$ and part (b) $z=7$.}
\end{figure}

\bigskip

\begin{figure}[hbt]
\begin{center}
\includegraphics[angle=-90,scale=0.46]{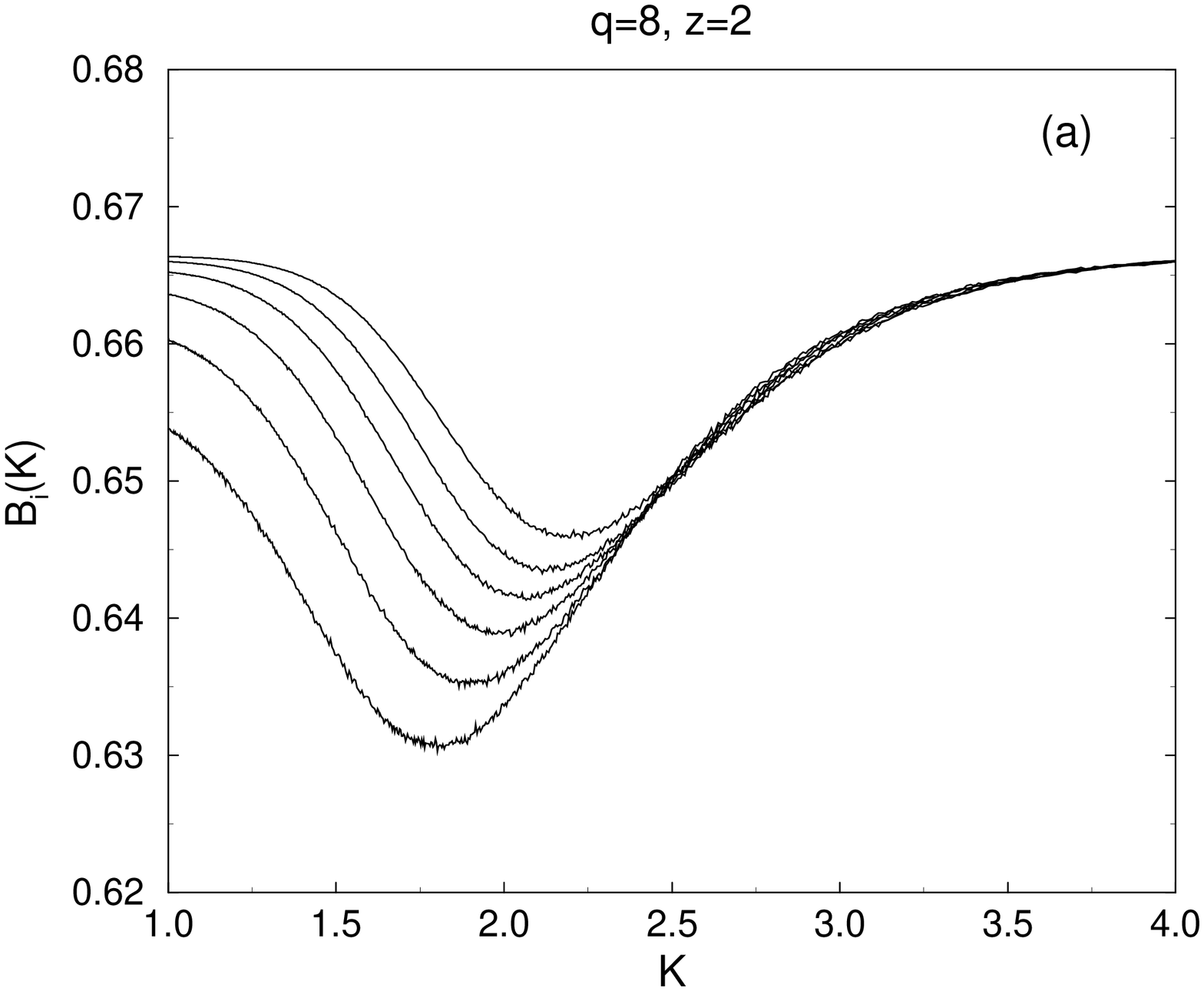}
\includegraphics[angle=-90,scale=0.46]{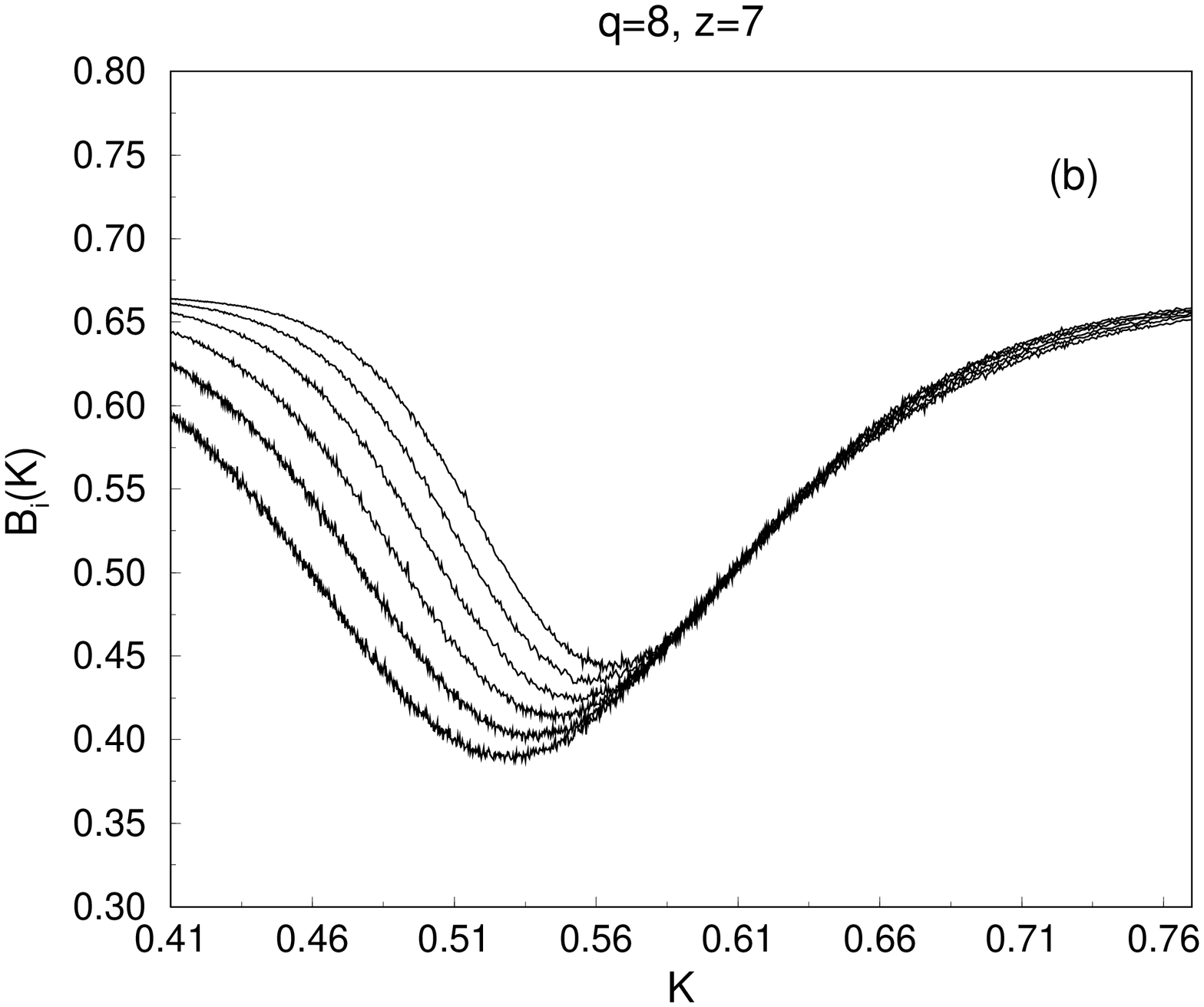}
\end{center}
\caption{Plot of the Binder cumulant $B_{i}(K)$ versus $K$ for $q=8$ and severals systems sizes ($N=250$, $500$, $1000$, $2000$, $4000$, and $8000$ ). In part (a) $z=2$ and part (b) $z=7$.
}
\end{figure}
\bigskip
 
\begin{figure}[hbt]
\begin{center}
\includegraphics[angle=-90,scale=0.46]{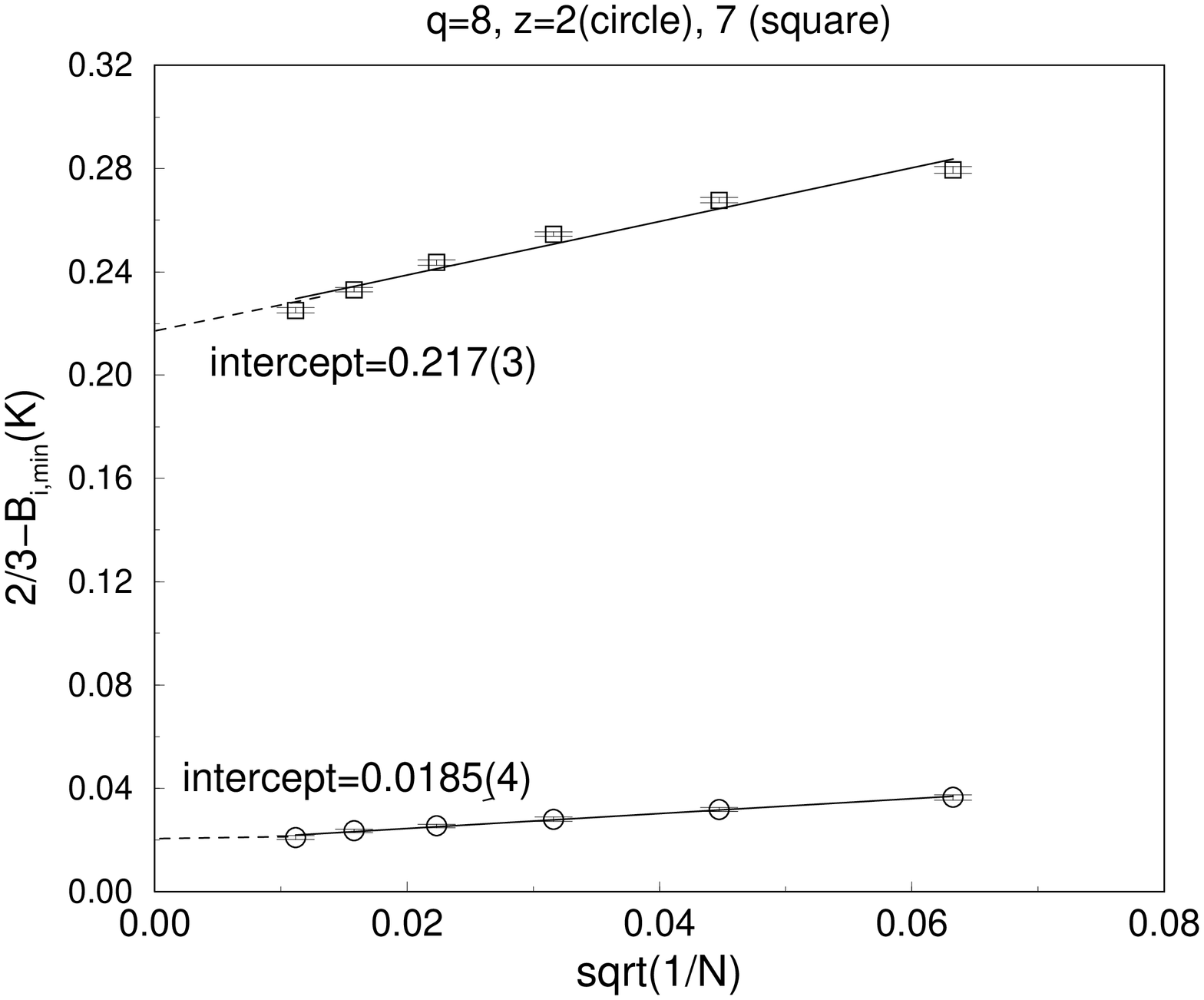}
\includegraphics[angle=-90,scale=0.46]{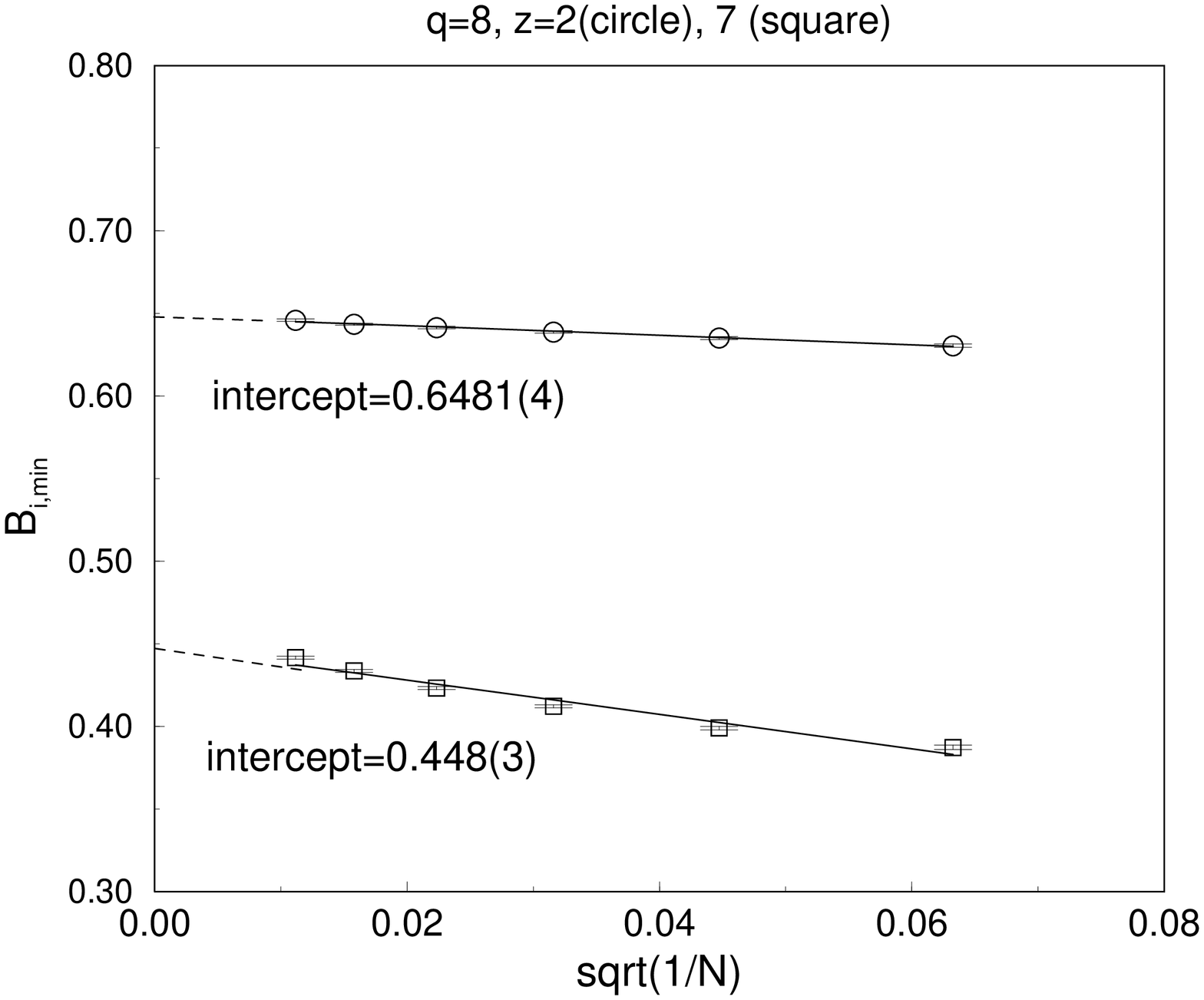}
\end{center}
\caption{Plot of the Binder parameter $2/3-B_{i}(K)$ versus $1/N$ for $z=2$ (circle) and $z=7$ (square),part (a) and  $B_{i,min}(K)$ versus $1/N$, part (b) for $q=8$.}
\end{figure}

\bigskip

\begin{figure}[hbt]
\begin{center}
\includegraphics[angle=-90,scale=0.46]{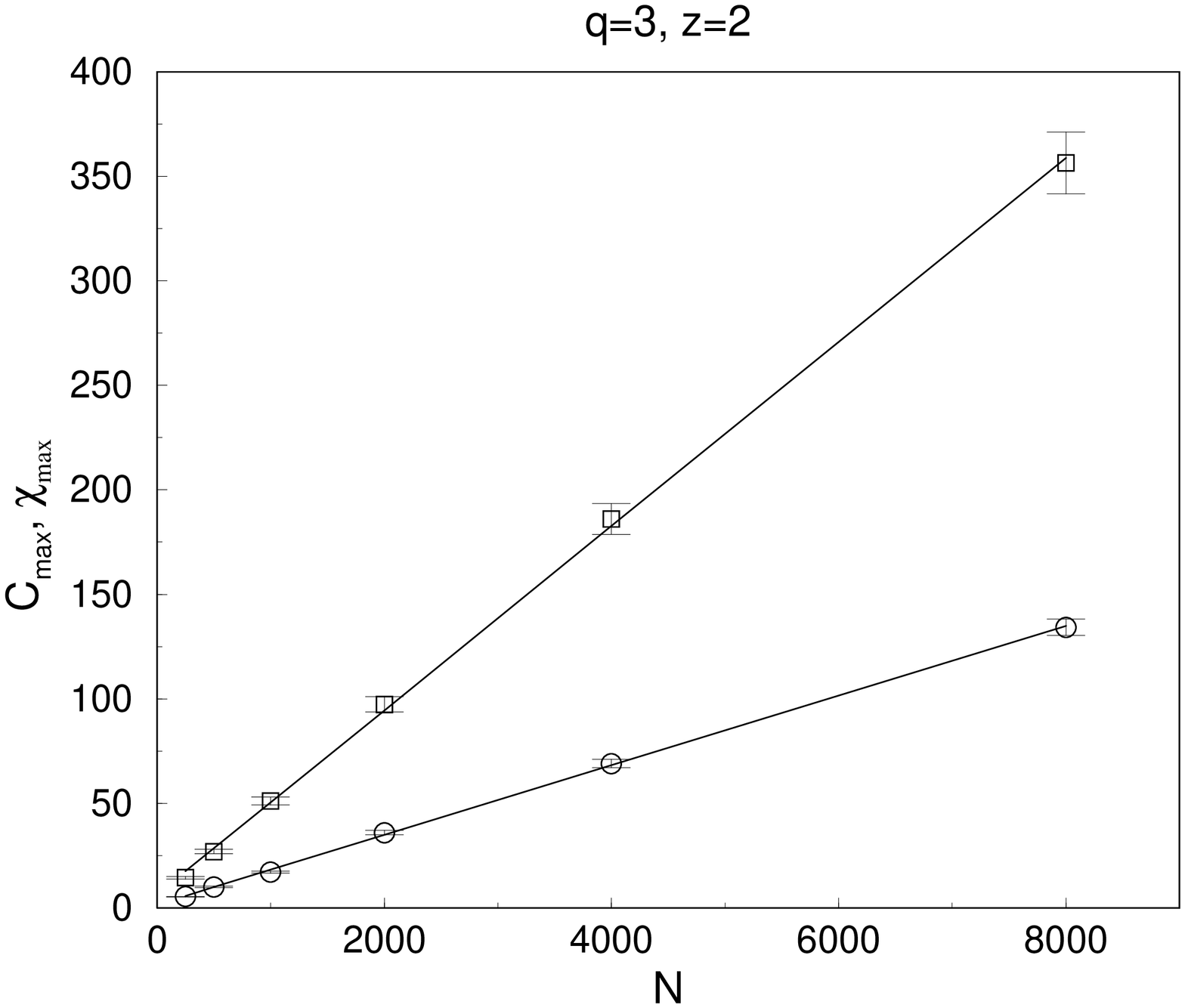}
\includegraphics[angle=-90,scale=0.46]{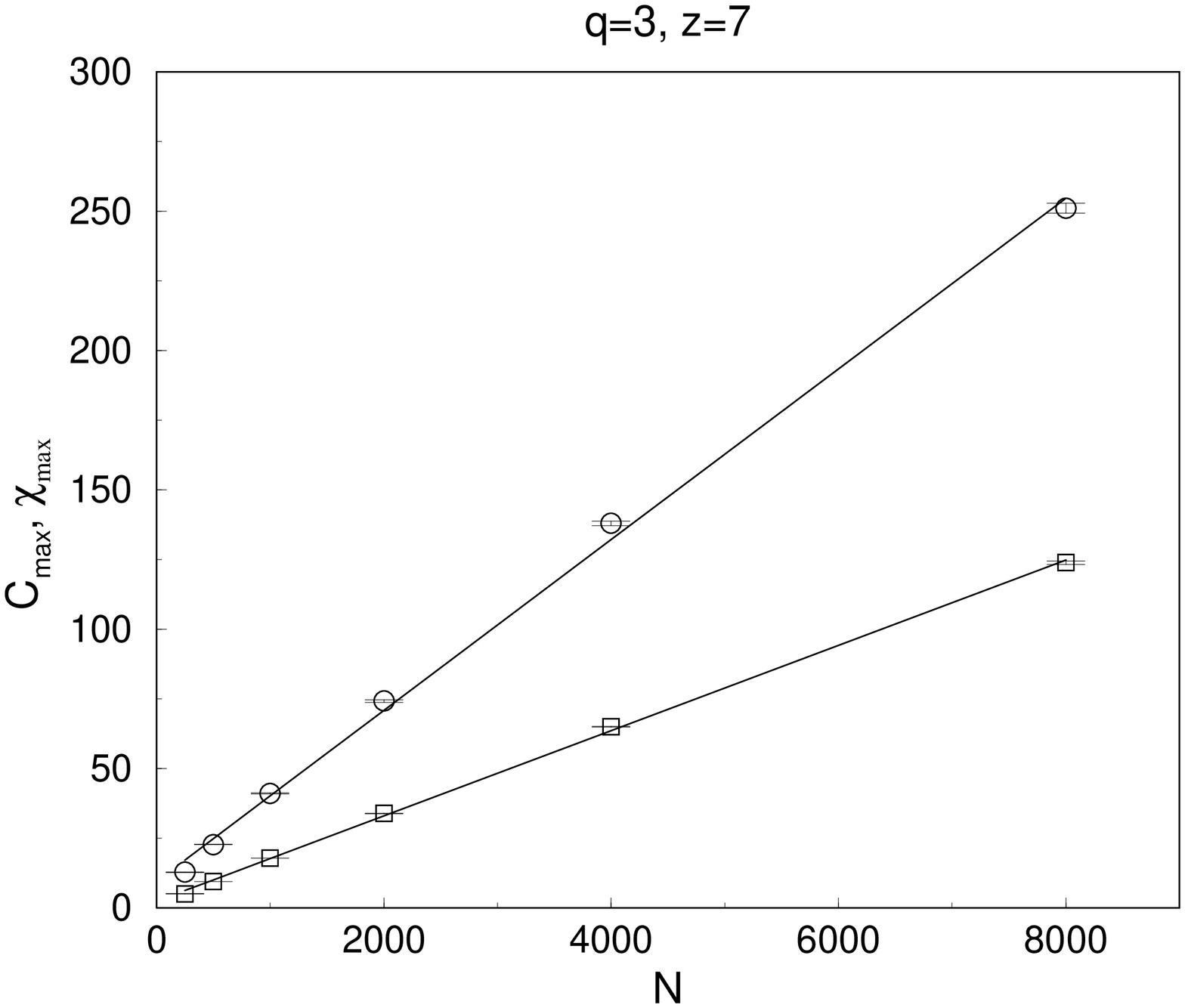}
\end{center}
\caption{Plot of the specific heat $C_{max}$ (circle) and susceptibility $\chi_{max}$ (square) versus $N$ for $q=3$. In part (a) $z=2$ and part (b) $z=7$.}
\end{figure}

\bigskip

\begin{figure}[hbt]
\begin{center}
\includegraphics[angle=-90,scale=0.46]{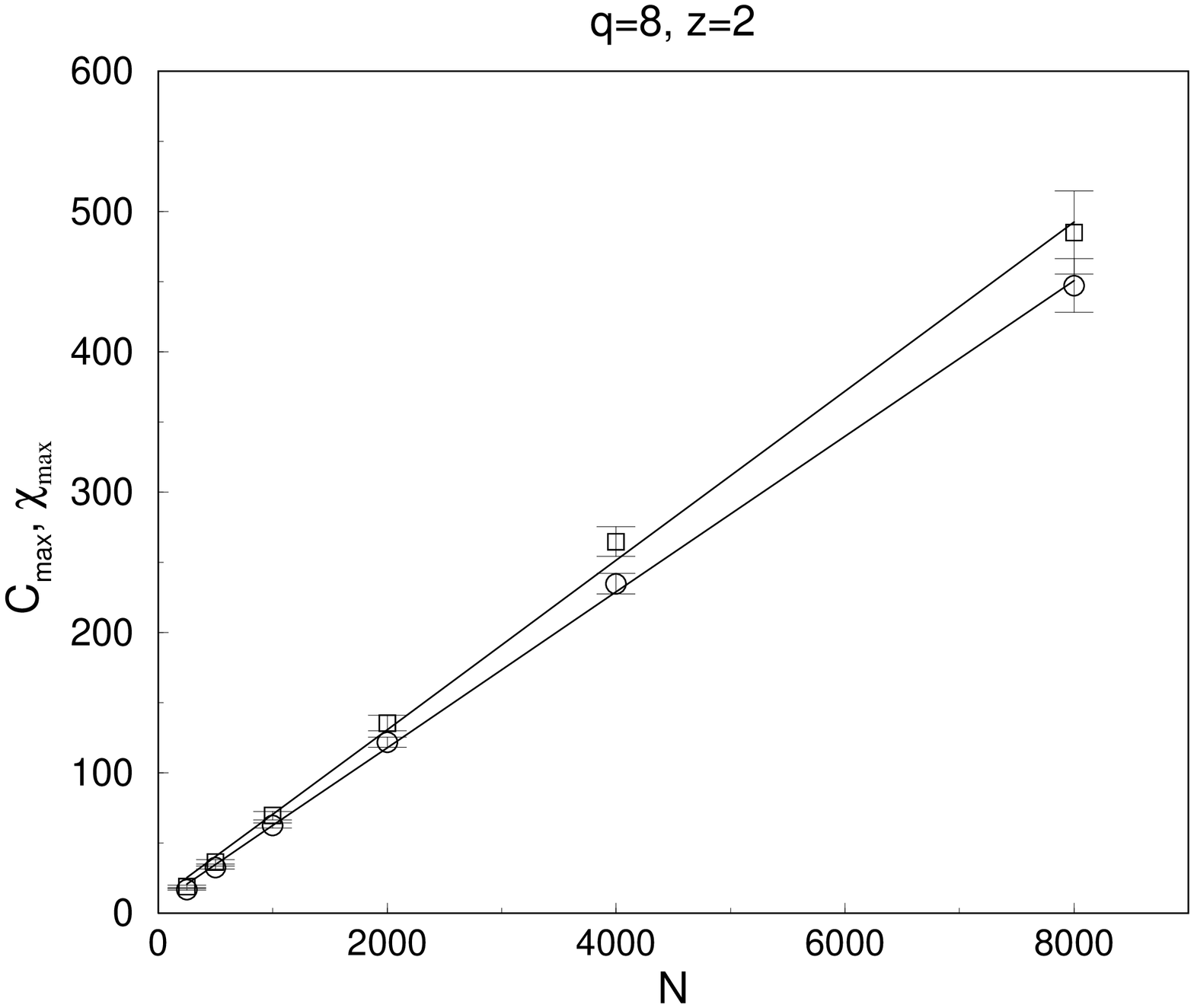}
\includegraphics[angle=-90,scale=0.46]{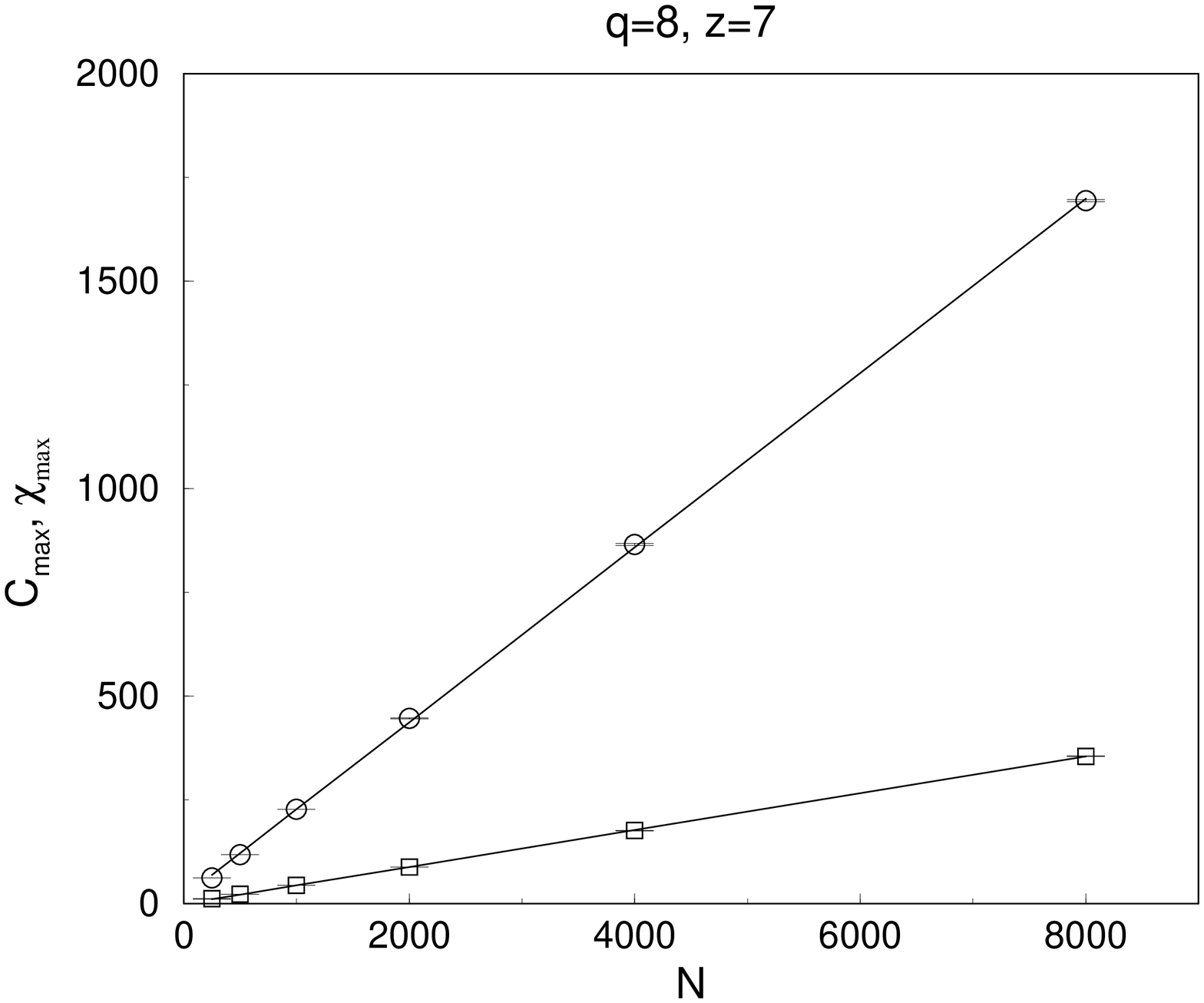}
\end{center}
\caption{Plot of the specific heat $C_{max}$ (circle) and susceptibility $\chi_{max}$ (square) versus $N$ for $q=8$. In part (a) $z=2$ and part (b) $z=7$}
\end{figure}

\bigskip

{\bf Model and Simulation}

We consider the Potts model with  $q=3$ and $8$ states, on directed 
Barab\'asi-Albert Networks, defined by a set of
spin variables ${\sigma}$ taking the values $1$, $2$, and $3$ for $q=3$ and
$\sigma=1$,...,$8$ for $q=8$ states situated on every site of a directed 
Barab\'asi-Albert Networks with $N$ sites. 
. 

The Potts interation energy is given by
\begin{equation}
E=-J\sum_{i}\sum_{k}\delta_{\sigma_{i}\sigma_{k}}
\end{equation} 
where $k$-sum runs over all nearest neighbors of $i$. In this network, each new site
added to the network selects $z$
already existing sites as neighbours influencing it; the newly
added spin does not influence these neighbours. 
To study the critical behavior of the model we define the variable $e=E/N$
and $m=(q\cdot \max_i [n_{i}]-N)/(q-1)$, where $n_{i}\le N$ denotes the number of spins with " orientation " $i=1,2,$ and $3$ and $i=1,2, ...,8$ for $q=3$
and $q=8$, respectively, in one network configuration.
>From the variable energy we can compute the average energy and specific heat and energetic fourth-order cumulant,
\begin{equation}
 u(K)=[<E>]_{av}/N,
\end{equation}
\begin{equation}
 C(K)=K^{2}N[<e^{2}>-<e>^{2}]_{av},
\end{equation}
\begin{equation}
 B_{i}(K)=[1-\frac{<e^{4}>}{3<e^{2}>^{2}}]_{av},
\end{equation}
where $K=J/k_BT$, with $J=1$, and $k_B$ is the Boltzmann constant. 
Similarly, we can derive from the magnetisation measurements
the average magnetisation, the susceptibility, and the magnetic
cumulants,
\begin{equation}
 m(K)=[<|m|>]_{av},
\end{equation}
\begin{equation}
 \chi(K)=KN[<m^{2}>-<|m|>^{2}]_{av},
\end{equation}
\begin{equation}
 U_{4}(K)=[1-\frac{<m^{4}>}{3<|m|>^{2}}]_{av}.
\end{equation}
where $<...>$ stands for a thermodynamics average and $[...]_{av}$ square brackets
for a averages over the 20 realizations. 

In the order to verify the order of the transition this model, we apply finite-size scaling (FSS) \cite{fss}. Initially we search for the minima of energetic fourth-order cumulant in eq. (4). This quantity gives a qualitative as well as a quantitative description of the order the transition \cite{mdk}. It is known \cite{janke} that this
parameter takes a minimun value $B_{i,min}$ at the effective transition temperature 
$T_{c}(N)$. One can show \cite{kb} that for a second-order transition $\lim_{N\to \infty}$
$(2/3-B_{i,min})=0$, even at $T_{c}$, while at a first-order transition the same limit measures the latent heat $|e_{+}-e_{-}|$:

\begin{equation}
\lim_{N\to \infty}(2/3-B_{i,min})=\frac{1}{3}\frac{(e_{+}-e_{-})^{2}(e_{+}+e_{-})^{2}}
{(e_{+}^{2}-e_{-}^{2})^{2}}.
\end{equation}

A more quantitative analysis can be carried out throught the FSS of the specific heat 
$C_{max}$, the susceptibility maxima $\chi_{max}$ and the minima of the Binder parameter $B_{i,min}$. If the hypothesis of a first-order phase transition is correct, we should then expect, for large systems sizes, an asymptotic FSS behavior of the form
\cite{wj,pbc},
\begin{equation}
C_{max}=a_{C} + b_{C}N +...
\end{equation}
\begin{equation}
\chi_{max}=a_{\chi} + b_{\chi}N +...
\end{equation}
\begin{equation}
B_{i,min}=a_{B_{i}} + b{B_{i}}N +...
\end{equation}

We have performed Monte Carlo simulation on directed Barab\'asi-Albert networks with
values of connectivity $z=2$ and $7$. For a given $z$, we used systems
of size $N=250$, $500$, $1000$, $2000$, $4000$, and $8000$ sites. We waited $10000$ Monte Carlo
steps (MCS) to make the system reach the steady state, and the time averages were
estimated from the next $ 10000$ MCS. In our simulations, one MCS is accomplished
after all the $N$ spins are updated. For all sets of parameters, we have generated
$20$ distinct networks, and have simulated $20$
independent runs for each distinct network.

\bigskip

{\bf Results and Discussion}

Our simulations, using the HeatBath algorithm, indicate that the model displays a first order phase transition.
In plot Fig. 1 we show the dependence of the magnetisation $M$ on the temperature $T$ for $q=3$ and several system sizes. Part (a) shows the  curves for $z=2$ of top to bottom of $N=250$ to $8000$, part (b)  the same as part (a) for $z=7$. In Fig. 2 we show the dependence of the Binder parameter $B_{i}(K)$ for connectivity $z=2$ and $7$ and various systems size.
Part (a) shows the  curves for $z=2$ of bottom to top of $N=250$ to $8000$, part (b)  the same as part (a) for $z=7$. The Binder parameter clearly goes to a value which is different from $2/3$. This is a sufficient condition to characterize a first-order transition.
In Fig. 3 we plot the Binder parameter $B_{i}$ versus $1/N$ for $z=2$ (circle) and $z=7$ (square), and severals systems sizes ($N=250, 500, 1000$, $2000$, $4000$, and $8000$), Part (a). We show the scaling of the Binder parameter minima, and again the first order phase transition is verified. 
The order of the transitions can be confirmed by plotting the values of $2/3-B_{i,min}$
again versus $1/N$. For a second-order transition the curves goes to zero as we increase the system size. Here, the quantity $2/3-B_{i,min}$ approaches a nonvanishing value in the
limit of small $1/N$, for $z=2$ as for $z=7$, see Fig. 3, part (b). In the
Figs. 4, 5, and 6 we show the same behavior the Figs. 1, 2, and 3 , respectively, but for $q=8$. As depicted in Figures 7 an 8, our results for scaling of the specific heat and susceptibility are consistent with equations (9,10). In Fig. 7 we show this behavior for $z=2$, part (a), and part (b), $z=7$ for $q=3$. The same occurs with Fig. 8, but for $q=8$.

\bigskip
 
\bigskip

{\bf Conclusion}
 
In conclusion, we have presented a very simple equilibrium model on
directed Barab\'asi-Albert network \cite{sumour,sumourss}. Different from
the spin 1/2 Ising model, in these networks,
the spin $q=3$ and $8$ Potts model presents a
the first-order phase transition which occurs in model with 
connectivity $z=2$ and $z=7$ here studied. These results disagree
with the results for Potts model on two-dimensional lattice \cite{kb}, where for $q\leq4$ the transitions is the order second  and the order first for $q>4$ .

  It is a pleasure to thank D. Stauffer for many suggestions and fruitful
discussions during the development this work and also for the revision of this paper.
I also acknowledge the Brazilian agency FAPEPI
(Teresina-Piau\'{\i}-Brasil) for  its financial support
and also the Fernando Whitaker for help in the support the system SGI Altix 1350 the computational park CENAPAD.UNICAMP-USP, SP-BRASIL.

\end{document}